\documentclass[aps,prc,twocolumn,superscriptaddress]{revtex4-2}

\usepackage{graphicx}

\usepackage{hyperref}
\hypersetup{colorlinks=true}

\usepackage{lineno}

\begin{document}


\title{Measurement of neutron-proton capture in the SNO+ water phase}



\author{\bf M.\,R.\,Anderson}
\affiliation{\it Queen's University, Department of Physics, Engineering Physics \& Astronomy, Kingston, ON K7L 3N6, Canada}
\author{\bf S.\,Andringa}
\affiliation{\it Laborat\'{o}rio de Instrumenta\c{c}\~{a}o e  F\'{\i}sica Experimental de Part\'{\i}culas (LIP), Av. Prof. Gama Pinto, 2, 1649-003, Lisboa, Portugal}
\author{\bf M.\,Askins}
\affiliation{\it University of California, Berkeley, Department of Physics, CA 94720, Berkeley, USA}
\affiliation{\it Lawrence Berkeley National Laboratory, 1 Cyclotron Road, Berkeley, CA 94720-8153, USA}
\author{\bf D.\,J.\,Auty}
\affiliation{\it University of Alberta, Department of Physics, 4-181 CCIS,  Edmonton, AB T6G 2E1, Canada}

\author{\bf N.\,Barros}
\affiliation{\it Laborat\'{o}rio de Instrumenta\c{c}\~{a}o e  F\'{\i}sica Experimental de Part\'{\i}culas (LIP), Av. Prof. Gama Pinto, 2, 1649-003, Lisboa, Portugal}
\affiliation{\it Universidade de Lisboa, Faculdade de Ci\^{e}ncias (FCUL), Departamento de F\'{\i}sica, Campo Grande, Edif\'{\i}cio C8, 1749-016 Lisboa, Portugal}
\author{\bf F.\,Bar\~{a}o}
\affiliation{\it Laborat\'{o}rio de Instrumenta\c{c}\~{a}o e  F\'{\i}sica Experimental de Part\'{\i}culas (LIP), Av. Prof. Gama Pinto, 2, 1649-003, Lisboa, Portugal}
\affiliation{\it Universidade de Lisboa, Instituto Superior T\'{e}cnico (IST), Departamento de F\'{\i}sica, Av. Rovisco Pais, 1049-001 Lisboa, Portugal}
\author{\bf R.\,Bayes}
\affiliation{\it Laurentian University, Department of Physics, 935 Ramsey Lake Road, Sudbury, ON P3E 2C6, Canada}
\author{\bf E.\,W.\,Beier}
\affiliation{\it University of Pennsylvania, Department of Physics \& Astronomy, 209 South 33rd Street, Philadelphia, PA 19104-6396, USA}
\author{\bf A.\,Bialek}
\affiliation{\it SNOLAB, Creighton Mine \#9, 1039 Regional Road 24, Sudbury, ON P3Y 1N2, Canada}
\author{\bf S.\,D.\,Biller}
\affiliation{\it University of Oxford, The Denys Wilkinson Building, Keble Road, Oxford, OX1 3RH, UK}
% ...
\collaboration{The SNO+ Collaboration}



\begin{abstract}
The SNO+ experiment collected data as a low-threshold water Cherenkov detector from September 2017 to July 2019. 
Measurements of the 2.2-MeV $\gamma$ produced by neutron capture on hydrogen have been made using an Am-Be calibration source, for which a large fraction of emitted neutrons are produced simultaneously with a 4.4-MeV $\gamma$. 
Analysis of the delayed coincidence between the 4.4-MeV $\gamma$ and the 2.2-MeV capture $\gamma$ revealed a neutron detection efficiency that is centered around 50\% and varies at the level of 1\% across the inner region of the detector, which to our knowledge is the highest efficiency achieved among pure water Cherenkov detectors.  In addition, the neutron capture time constant was measured and converted to a thermal neutron-proton capture cross section of $336.3^{+1.2}_{-1.5}$~mb.  
\end{abstract}


\maketitle

\section{Introduction}
\label{sec:intro}
Detecting neutron captures is important for the identification of both signals and backgrounds in low-energy nuclear and particle physics experiments.  
Inverse beta decays ($\overline{\nu}_e + p \rightarrow e^+ + n$)  have been used in the first detection of antineutrinos, in reactor antineutrino oscillation measurements, in the discovery of geo-antineutrinos, and in measurements of, and searches for, supernova antineutrinos. 
Backgrounds with associated neutrons include $\alpha$-$n$ reactions from radioactive isotopes, $\beta$-$n$ decays of cosmogenically-produced isotopes, and cosmogenically-induced spallation neutrons.  
In many cases, radiative capture signals provide the most reliable means to identify the neutron, and in general their delay in relation to the neutron production time provides a clear signature.  

In pure water- or liquid scintillator-based neutrino experiments, a neutron will capture on a hydrogen nucleus with a time constant of about 0.2~ms, emitting a 2.2-MeV $\gamma$ with the production of $^2$H.  
This signal is used, for example, in Refs.~\cite{Eguchi:2002dm, Bellini:2010hy, Zhang:2013tua}. 
Other experiments have used nuclei with relatively large neutron capture cross sections, such as Cl~\cite{Boger:1999bb}, Gd~\cite{An:2015qga,Allemandou:2018vwb,Back:2019aqi}, Li~\cite{Ashenfelter:2018zdm} or B~\cite{Agnes:2015qyz}, to shorten the neutron capture time and often produce more distinctive capture signals.  

SNO+ has acquired data for two years as a kiloton-scale pure water Cherenkov detector.  
With a low trigger threshold, SNO+ has a relatively high efficiency for detecting 2.2-MeV $\gamma$'s in pure water.
Using a deployed Am-Be calibration source, which emits a 4.4-MeV $\gamma$ for a large fraction of emitted neutrons, both the detection efficiency and capture time of neutrons were measured.  

The detector and trigger scheme, and the Am-Be source and its deployments, are described in Sections~\ref{sec:detector} and \ref{sec:source}, respectively.  
Section~\ref{sec:analysis} describes the analysis of the Am-Be data to determine the neutron detection efficiency across the detector volume and measure the neutron capture time constant in water.  
In Section~\ref{sec:CS}, the capture time constant is converted to a thermal neutron-proton capture cross section.

\section{SNO+ detector and trigger}
\label{sec:detector}
SNO+ is a multipurpose neutrino experiment with the primary goal of searching for neutrinoless double beta decay~\cite{Andringa:2015tza}.  Three operational phases using different target materials are scheduled: water, scintillator, and Te-loaded scintillator.  The completed water phase was required to calibrate detector components and measure the intrinsic levels of radioactivity in the detector materials.  
With an initial data set, SNO+ measured $^8$B solar neutrinos with low backgrounds~\cite{Anderson:2018ukb} and set world-leading limits on invisible modes of (di)nucleon decay~\cite{Anderson:2018byx}.  
The ongoing scintillator phase is required to characterize the scintillator, and will be used to make measurements of reactor and geo antineutrinos, and potentially lower energy solar neutrinos.  
The third phase will be dedicated to a search for the neutrinoless double beta decay of $^{130}$Te, while continuing measurements of antineutrinos.  

Most of the infrastructure of the experiment is inherited from the Sudbury Neutrino Observatory (SNO), which used heavy water (D$_2$O) as a target~\cite{Boger:1999bb}.
The experiment is located 2.0~km underground at SNOLAB, in Sudbury, Ontario, Canada.  
The target liquid is contained within a 5-cm thick acrylic vessel (AV) with a radius of 6.0~m, which is submerged in water.  
Surrounding the AV, a geodesic structure with a radius of 8.9~m supports more than 9300 Hamamatsu R1408 photomultiplier tubes (PMTs) that face inward.  
The PMTs are each equipped with a light concentrator, yielding an effective optical coverage of approximately 54\%.  
A 6.8-m tall acrylic cylinder of 0.75~m radius extends from the top of the AV, providing necessary detector access, such as the deployment of calibration sources. 

The primary detector trigger is based on a sum of analog, fixed-current pulses from individual PMT channels.
Every PMT signal above its channel's threshold results in the production of a fixed-current pulse of 89~ns width.
These pulses are continuously summed across all inward-facing PMTs, and this sum is discriminated against an adjustable trigger threshold.
The behavior of the trigger system around threshold is governed primarily by the finite rise time of the pulses, the intrinsic noise on the analog sum, and shifting in the channel baselines.  The latter is the primary limit to further lowering the detector trigger threshold.  
Each of the above characteristics was measured during data acquisition and is modeled in the SNO+ simulation.

\begin{figure}[tbp]\centering
\includegraphics[width=1\linewidth]{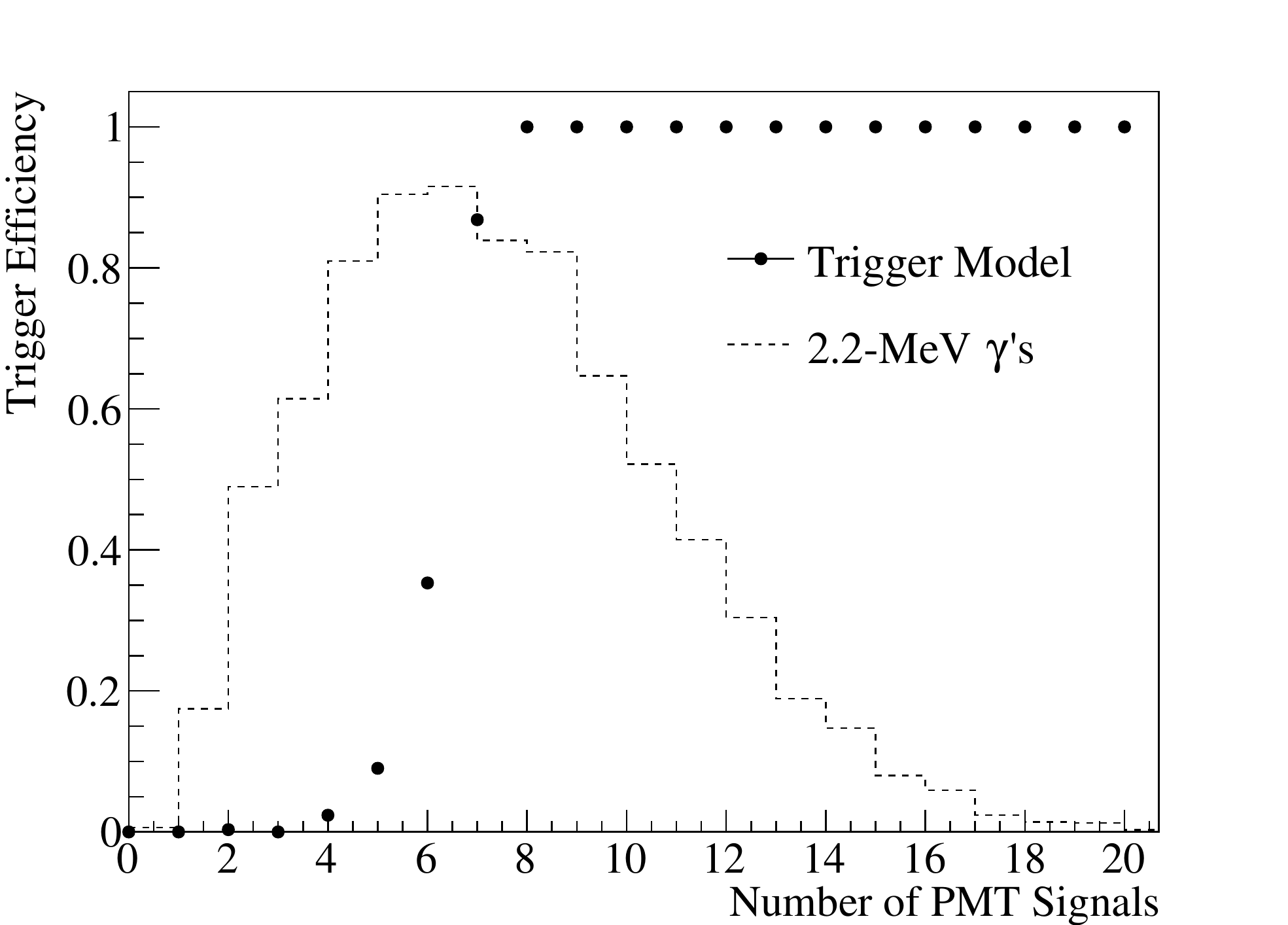}
\caption{Simulated trigger efficiency as a function of the number of PMT signals that contribute to the trigger (solid points), and the predicted distribution of the number of PMT signals from 2.2-MeV $\gamma$'s inside the AV (dashed line - arbitrary normalization).}
\label{fig:trigEffCurves}
\end{figure}

For the data analyzed in this article, the trigger threshold was set to 7.0 pulses, which corresponds to approximately 1.4~MeV for an electron at the center of the detector.  
Figure~\ref{fig:trigEffCurves} shows the probability to trigger as a function of the number of PMT signals that contribute to the trigger, together with a distribution of the number of PMT signals, both from a simulation of 2.2-MeV $\gamma$'s uniformly distributed inside the AV.  
The simulation suggests a trigger efficiency of 100\% for $\ge$~8 PMT signals.  
Convolving the distribution of the number of PMT signals with the trigger efficiency gives a total efficiency to trigger on 2.2-MeV $\gamma$'s around 50\%.   

\section{A\lowercase{m}-B\lowercase{e} source and deployment}
\label{sec:source}

In the Am-Be calibration source, $^{241}$Am nuclei undergo $\alpha$-decay with a half-life of 432 years and $^9$Be target nuclei absorb the emitted $\alpha$’s, producing a $^{12}$C nucleus and a neutron. The majority of neutrons thermalize in the detector and capture on hydrogen, emitting a 2.2-MeV $\gamma$ with the production of $^2$H. The $^{12}$C nucleus is produced in an excited state approximately 60\% of the time, from which it immediately de-excites, emitting a 4.4-MeV $\gamma$. Additional $\gamma$ emission from other excited states of carbon or from excitations of the oxygen in the water by neutrons from the source can contribute with smaller numbers of prompt events at higher energies. The coincidence between any of these prompt signals, namely the 4.4-MeV $\gamma$, and the delayed 2.2-MeV $\gamma$ provides a distinctive signature for identifying the Am-Be source neutrons.

The Am-Be source used in SNO+ is a powder source produced in 2005 and since stored at SNOLAB.  It is estimated to have had a rate of ($67.4\pm0.7$)~$n$/s at the time of deployment (2018), based on a measurement of its neutron rate in 2006~\cite{Loach:2008msa}.  
The source came doubly encapsulated in a stainless steel cylinder of 0.8-cm diameter and 1.0-cm height; 
but was further shielded with black Delrin\textsuperscript{\textregistered}~\cite{Delrin} thermoplastic encapsulation before use in SNO and again in SNO+, for compatibility with the deployment system and due to the cleanliness requirements of both experiments.  
The fully encapsulated source is a cylinder measuring approximately 6~cm in diameter and 8~cm in height.
Simulations of the Am-Be source and its encapsulation were performed to evaluate systematic effects, and are described in Sections~\ref{sec:syst} and \ref{sec:CSsyst}.

\begin{figure}[tbp]\centering
\includegraphics[width=\linewidth]{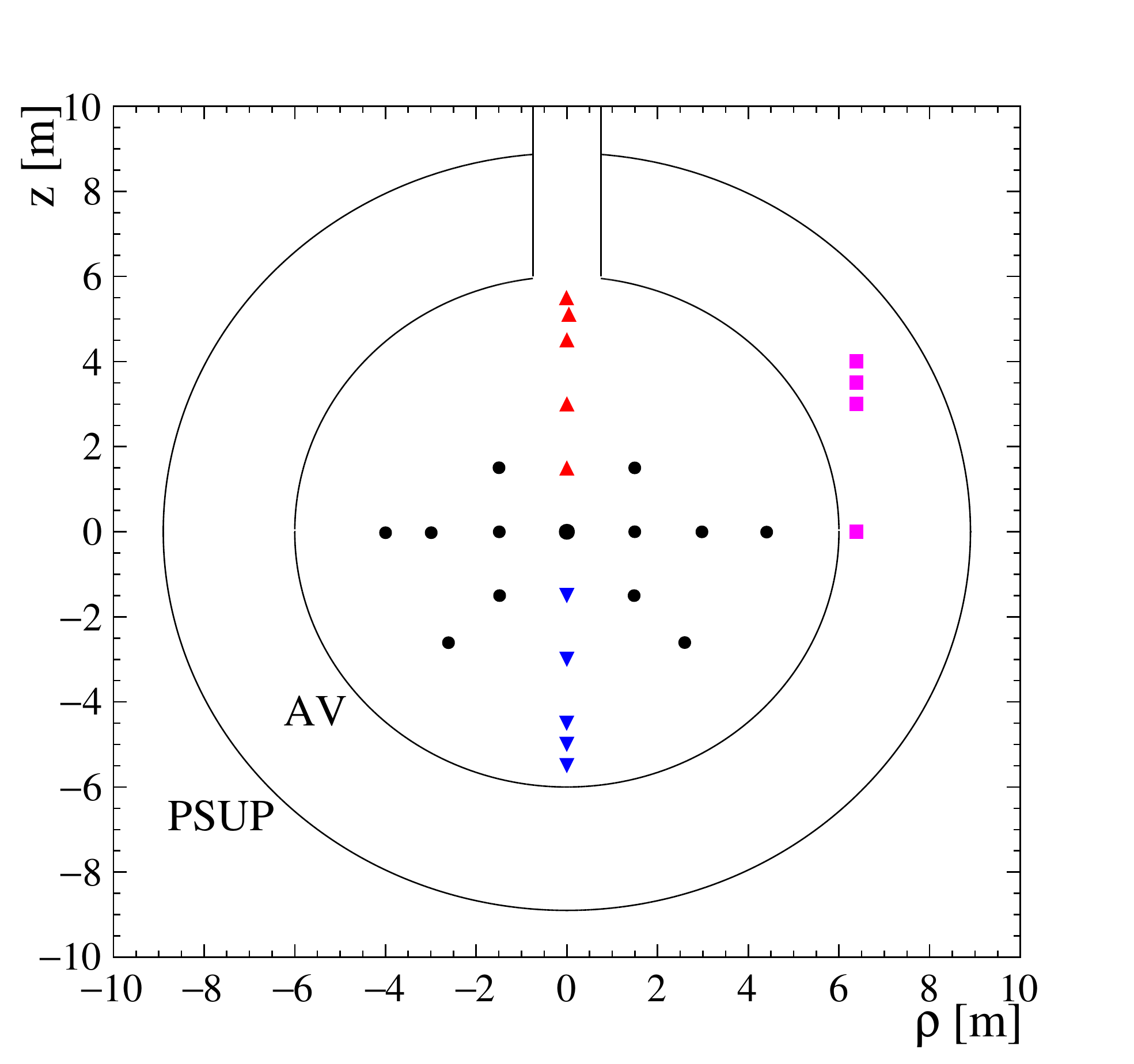}
\caption{SNO+ detector schematic showing the deployment positions of the Am-Be source that were used in the current analysis.  Data were collected along the central vertical axis ($\rho\equiv\sqrt{x^2+y^2}=0$) at $z >$ 0 (red) and $z <$ 0 (blue).  All other positions inside the AV are black and those outside the AV are magenta.   The AV has a radius of 6~m and the PMT support structure (PSUP) has a radius of 8.9~m.}
\label{fig:deployPos}
\end{figure}

The Am-Be source was deployed with a source manipulator system~\cite{Boger:1999bb} to 23 positions inside the AV in January 2018. Three hours of data were acquired at the center, and another 17 hours were used to scan a horizontal axis and the vertical axis.  
In June 2018, the source was deployed in the external water region along 13 vertically-aligned positions between the AV and the PMTs (only 4 were used in this analysis in order to preserve consistent trigger settings across all data).  
These positions are illustrated in Fig.~\ref{fig:deployPos} with a color scheme that is also used in Figs.~\ref{fig:NhitRun} and \ref{fig:eff}.

\section{Neutron capture analysis}
\label{sec:analysis}
The calibration data were collected with the same detector and trigger settings as used for normal data acquisition~\cite{Anderson:2018byx}. The same data cleaning procedures were applied to reduce instrumental effects, including the rejection of events that came in bursts, that followed within 3~$\mu$s after a trigger, and that followed within 20~s after the passage of a muon.  
Additionally, signals from PMTs were rejected if they arose from cross-talk at the front-end electronics or from PMTs that were insufficiently well calibrated.  The remaining PMT signals are referred to as $hits$. 
Contributions of events from $\alpha$-$n$ and $\beta$-$n$ processes were negligible relative to the source rate~\cite{Anderson:2018byx}, 
meaning that the dominant backgrounds in the measurements were accidental coincidences of the source $\gamma$'s and natural radioactivity in the detector.  

\begin{figure}[t]\centering
\includegraphics[width=1\linewidth]{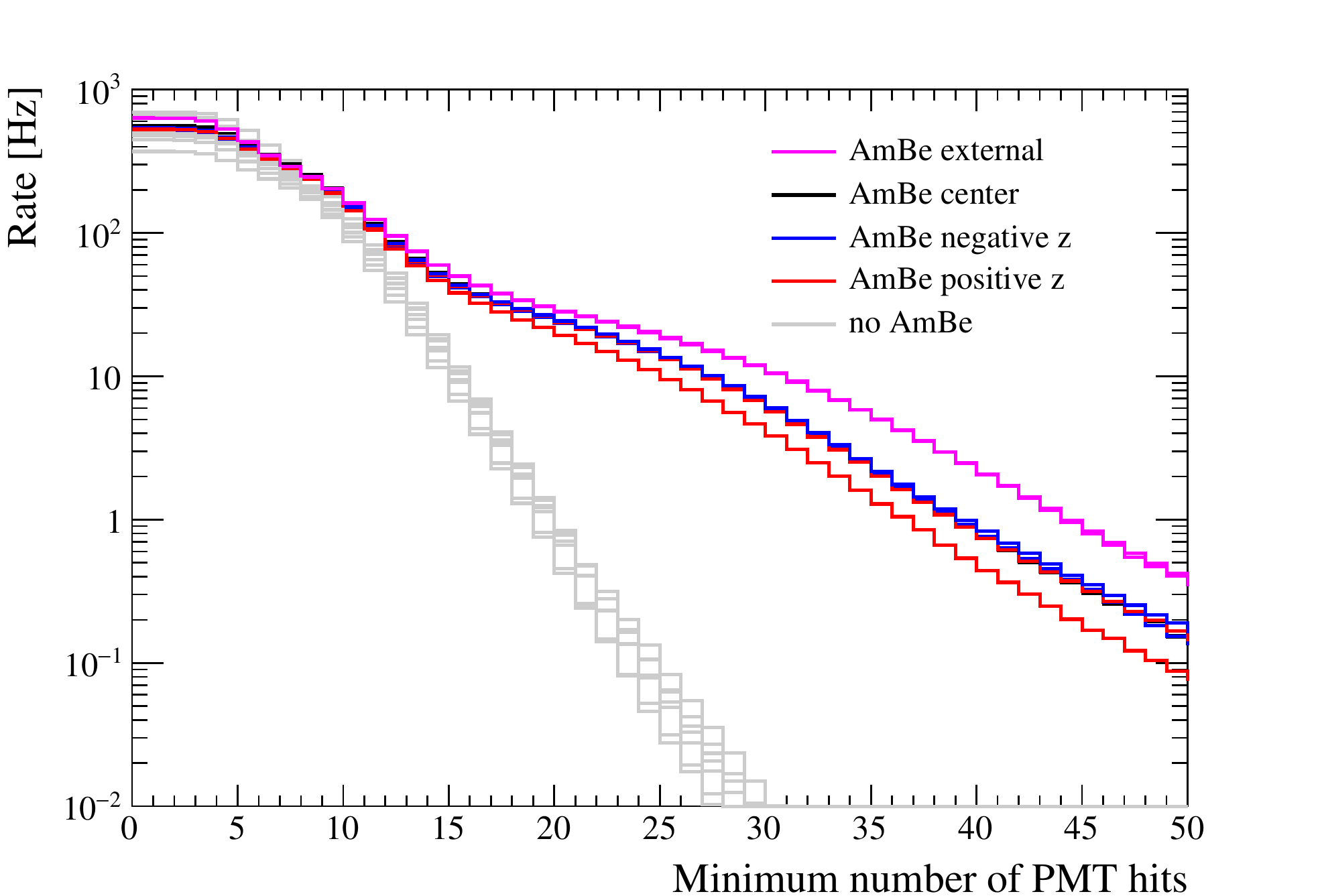}
\caption{Event rates vs. minimum number of PMT hits for Am-Be data collected along the central vertical axis, at $z$ = +1.5~m and +5.0~m (red), $z$ = 0~m (black), and $z$ = -1.5~m and -5~m (blue); and external to the AV at a radius of 6.4~m (magenta).  Several one hour periods just before and after Am-Be source deployment are shown in gray.}
\label{fig:NhitRun}
\end{figure}

Figure~\ref{fig:NhitRun} shows the event rate as a function of a minimum number of hits for data taken when the source was placed at different example positions, and when no source was deployed.  
The distributions just before and after deployment are shown in gray and exhibit similar behavior.  When the source was deployed, the rate of events with a high number of PMT hits increased, similarly so for different positions.  A slightly lower rate resulted when the source was placed closer to the top of the detector, where there are fewer PMTs due to the presence of the AV neck.  
Events due to the 4.4-MeV $\gamma$ from the source are clearly seen above about 15 PMT hits.  In contrast, the rate below 15 hits is dominated by detector backgrounds, obscuring the 2.2-MeV $\gamma$'s from neutron capture, but the average rate increase is compatible with that expected from the source activity.

\subsection{Analysis of coincidences}
\label{sec:minimalistic}
The Am-Be calibration data were analyzed using the difference between the trigger times of consecutive events, excluding only those with a number of PMT hits below threshold.

\begin{figure}[tbp]\centering
\includegraphics[width=1\linewidth]{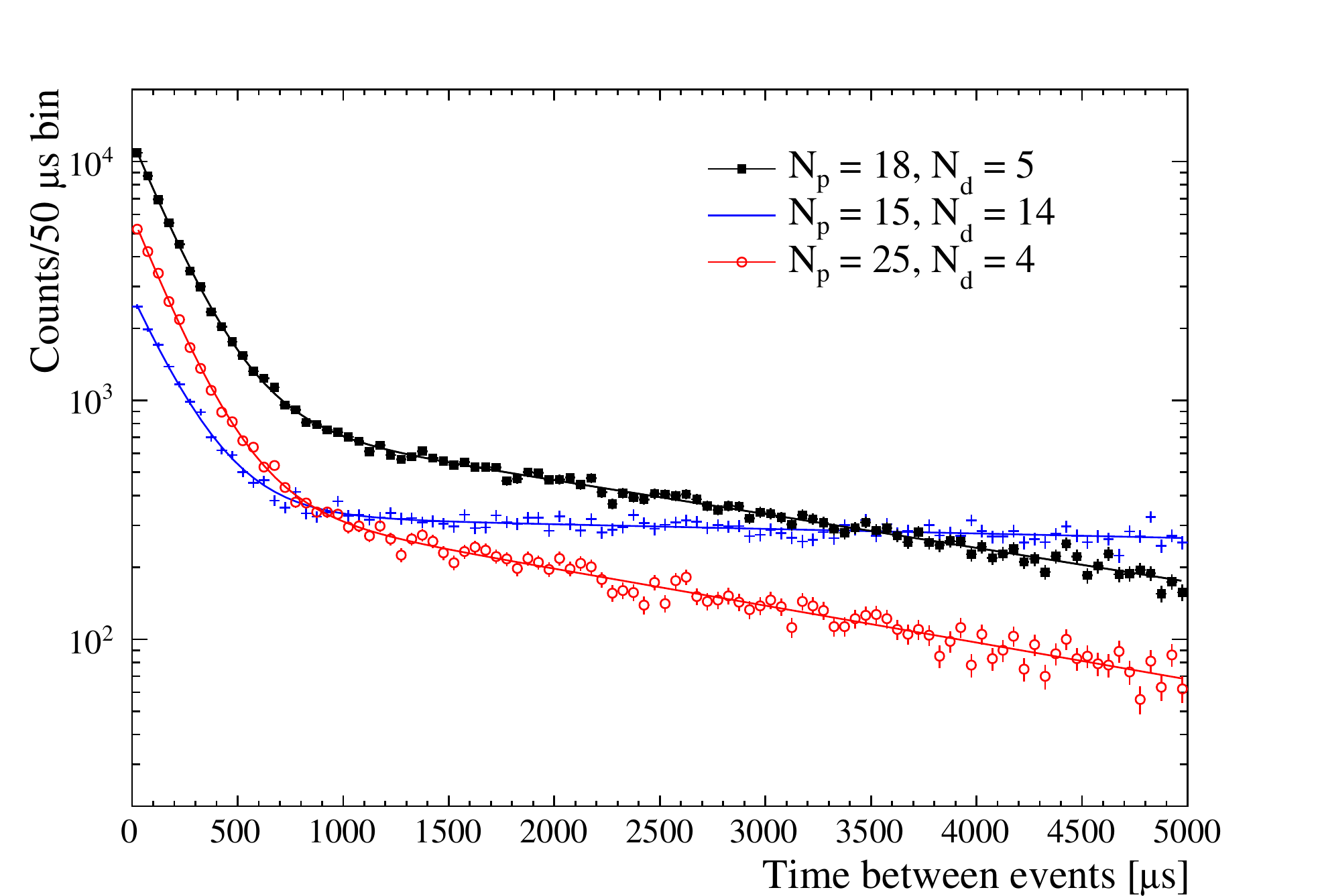}
\caption{Time between events for an Am-Be source deployment at the detector center.  Three different event selections are shown, with different thresholds for the PMT hits of prompt ($N_p$) and delayed ($N_d$) events.  The data exhibit two exponential decays, which are present in the fit function; Eq.~(\ref{eq:StatFit}).}
\label{fig:3fits}
\end{figure}

Data acquired with the Am-Be calibration source at the center of the detector are presented in Fig.~\ref{fig:3fits} with three basic event selections, which differ only in hit thresholds.  
The distribution of time between an event with at least 18 hits and the next event with at least 5 hits is shown in black.  Two exponentials are apparent: a faster one originating from the delayed neutron capture, and a slower one corresponding to random coincidences (extending well beyond the 5 ms shown).
The contribution from either exponential depends on the efficiency to detect the neutron and on the purity of selecting the correlated prompt 4.4-MeV $\gamma$. 
As a result, the distribution changes with the prompt and delayed hit thresholds, $N_p$ and $N_d$, as demonstrated by the other two spectra in Fig.~\ref{fig:3fits}.  

The distribution of time between events was fitted at each source deployment position with 
\begin{eqnarray}
\frac{dN}{dt} = T \cdot R_p [ PE\cdot(\lambda+R_d) \exp(-(\lambda+R_d)t) \nonumber\\
+ (1-PE)\cdot R_d \exp(-R_dt) ],
\label{eq:StatFit}
\end{eqnarray}
where $T$ is the data acquisition time, $R_p$ and $R_d$ are the rates of single events with a number of hits greater than or equal to $N_p$ and $N_d$, respectively, and $\lambda\equiv\tau^{-1}$ is the inverse of the neutron capture time constant.  
The parameter $PE$ is the product of the purity of 4.4-MeV $\gamma$'s among prompt events, $P$, and the efficiency to detect the neutron capture signal, $E$.  
Random coincidences occur when the prompt event is not the 4.4-MeV $\gamma$ or when the 2.2-MeV $\gamma$ is not detected, which implies a rate proportional to $(1-P)+P(1-E)=1-PE$.  For the cases in which a neutron capture is detected after a prompt signal, the coincidence rate is represented by $\lambda+R_d$, which accounts for when an uncorrelated event is detected before the neutron capture.  Equation~(\ref{eq:StatFit}) neglects the rare cases of two consecutive prompt-like events.  This creates a bias of order 0.1\% in $\lambda$, which is accounted for in the correction described in Section~\ref{sec:syst}.  

A series of fits was performed for each source position, with three free parameters ($PE, \lambda, R_d$).  
First, data were selected with thresholds of $N_p=15$ and $N_d=4$, and then each threshold was scanned individually.  

\begin{figure}[tbp]\centering
\includegraphics[width=1\linewidth]{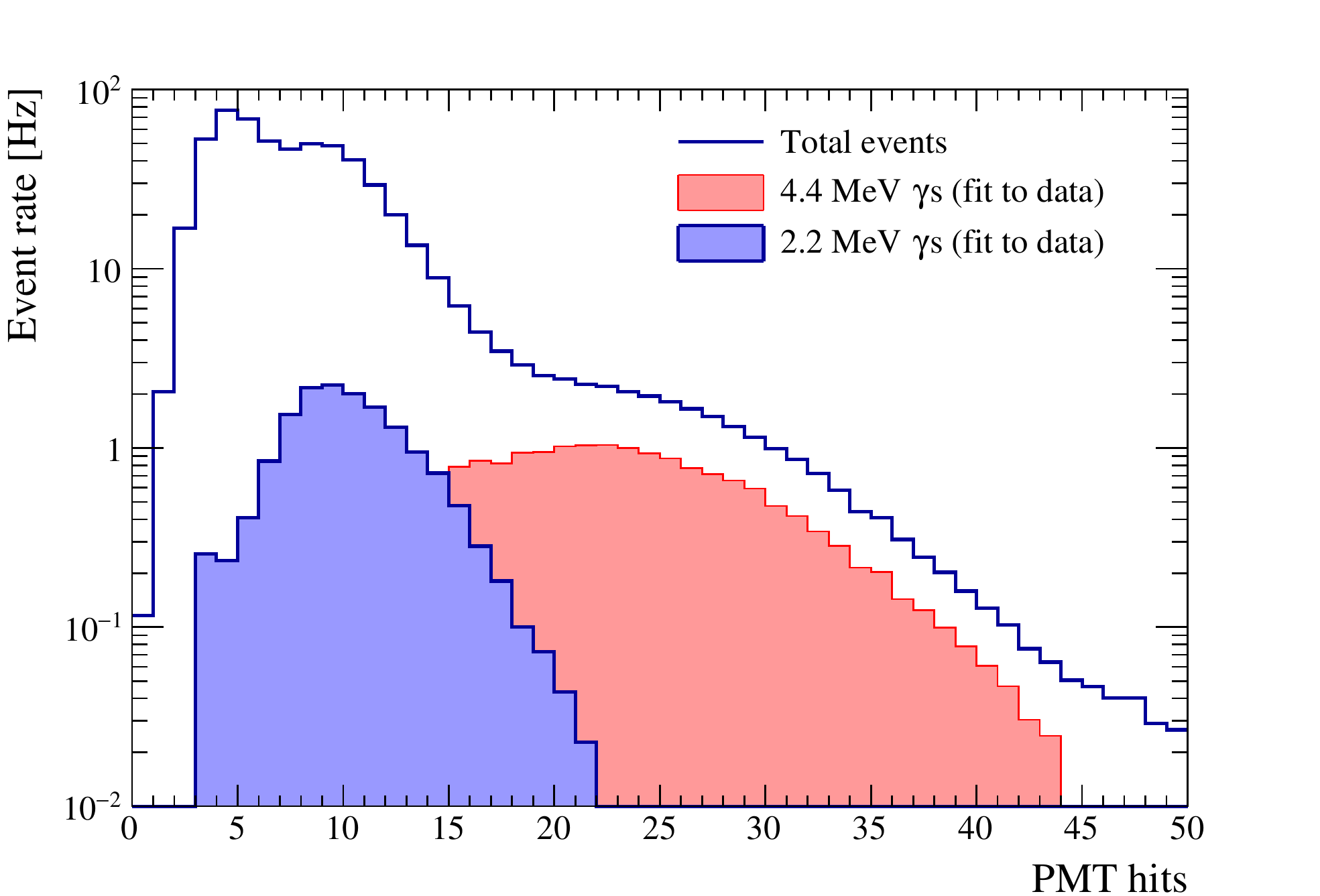}
\caption{Event rate distributions of PMT hits for the 2.2-MeV and 4.4-MeV $\gamma$'s, constructed from the time coincidence fits. Distributions are not stacked. The distribution from all events acquired with the Am-Be source at the center of the detector is also shown.
}\label{fig:NhitCenter}
\end{figure}

Calculating the rate of Am-Be coincidence events ($R_pPE$) as a function of $N_p$ and $N_d$ allows the construction of the PMT hit distributions of the prompt and delayed $\gamma$'s, as shown in Fig.~\ref{fig:NhitCenter} for a central deployment.
Most of the 4.4-MeV $\gamma$ distribution is in the region where SNO+ has 100\% trigger efficiency, and a majority of the 2.2-MeV $\gamma$ distribution below this region can be constructed. 

The maximum value of $PE$ was obtained at $N_p=25$ (for high $P$) and $N_d=4$ (for high $E$). 
The selection at $N_p=25$ keeps around half of the prompt events from the source, which is reflected in a decrease of the fit normalization parameter $R_p$. The purity for $N_p=25$ was found to be (99.62 $\pm$ 0.15)\% by comparing event rates from when the source was at the center to those when no source was deployed, all of which are shown in Fig.~\ref{fig:NhitRun}.

\begin{figure}[htbp]\centering
\includegraphics[width=1.08\linewidth]{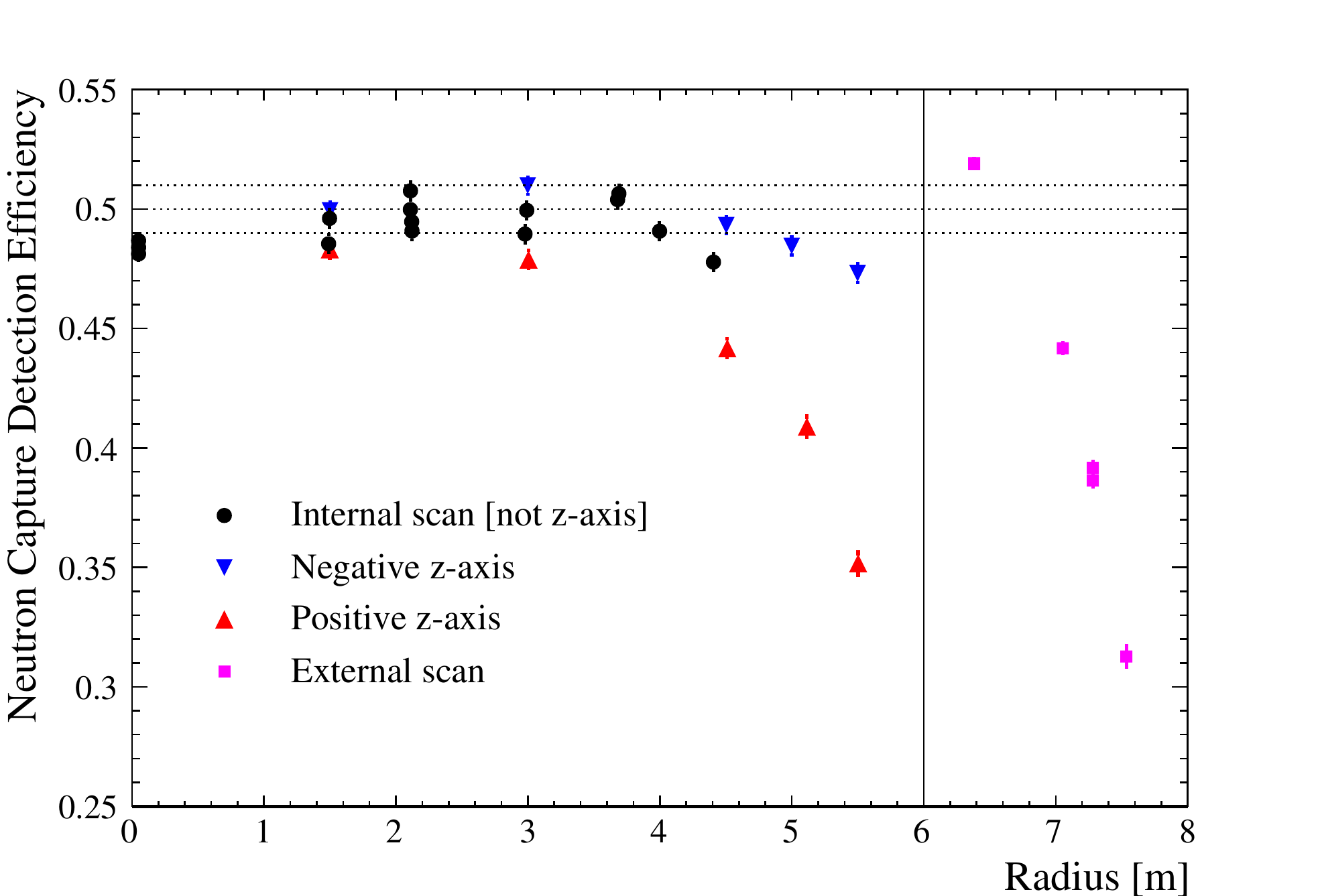}
\caption{Detection efficiency for the 2.2-MeV $\gamma$ from neutron capture obtained at each source position, as a function of radius. Error bars are statistical only.  Points along the central vertical axis of the detector are separated into positive (red) and negative (blue) $z$ positions.}\label{fig:eff} 
\end{figure}

Figure~\ref{fig:eff} shows the fitted detection efficiency without the small correction for purity (i.e., $PE$) as a function of the radial position in the detector. 
For reference, three hours of data collected at the center of the detector yielded $PE = (48.26\pm0.15)\%$ and $E=(48.44\pm0.17)\%$.  
The efficiency for detecting a 2.2-MeV $\gamma$ is around 50\% for radii up to about 4~m, with a variation at the level of $\pm$1\%. At heights $z>$~4~m, the AV neck and associated absence of PMTs introduce a significant vertical asymmetry: the efficiency is 47\% at $z$~= $-$5.5~m and 35\% at $z$~= $+$5.5~m. 
For the external deployments, the source was placed at radii between 6.4~m and 7.5~m (see Fig.~\ref{fig:deployPos}). 
The efficiency just outside the AV is higher than that at all internal positions due to the optical absorption and reflection of the AV.  
The efficiency drops quickly as the source approaches the PMTs; however, it is still above 30\% out to a radius of 7.5~m.  
If used, the AV-external volume between 6.0 to 7.5~m would almost double the fiducial volume available for low-energy analyses.
All features in Fig.~\ref{fig:eff} are present in the SNO+ simulation.  

The total uncertainty in the neutron capture time constant was minimized by setting the thresholds to $N_p=18$ and $N_d=5$. 
The results obtained for all AV-internal source positions were combined in an uncertainty-weighted average, resulting in $\tau = \lambda^{-1} = (207.03 \pm 0.42)~\mu$s.

\subsubsection{Higher-purity analysis of coincidences}
Another analysis of coincidences was applied to the three hours of data collected with the Am-Be source at the center of the detector.  It uses position reconstruction and requires that the two events be spatially coincident.  As a result, accidental backgrounds are reduced, but there are additional systematics
associated with reconstruction and detector modeling.  
The approach was used by Super-Kamiokande~\cite{Zhang:2013tua, Super-Kamiokande:2015xra} to design a dedicated trigger of delayed neutron captures following very high-energy events.  It can also be useful in studies of antineutrinos in SNO+.  

Following Ref.~\cite{Zhang:2013tua}, the analysis made use of the reconstructed position of the prompt event and the hit times of the delayed event.  
These hit times were tested against the hypothesis that the delayed event occurred at the same position as the prompt event.  
Each hit time was corrected for the time-of-flight calculated from the reconstructed position of the prompt event, and then only coincidences in which at least four of the hit times occurred within a sliding 12-ns window were accepted.  

\begin{figure}[tbp]\centering
    \includegraphics[width=1\linewidth]{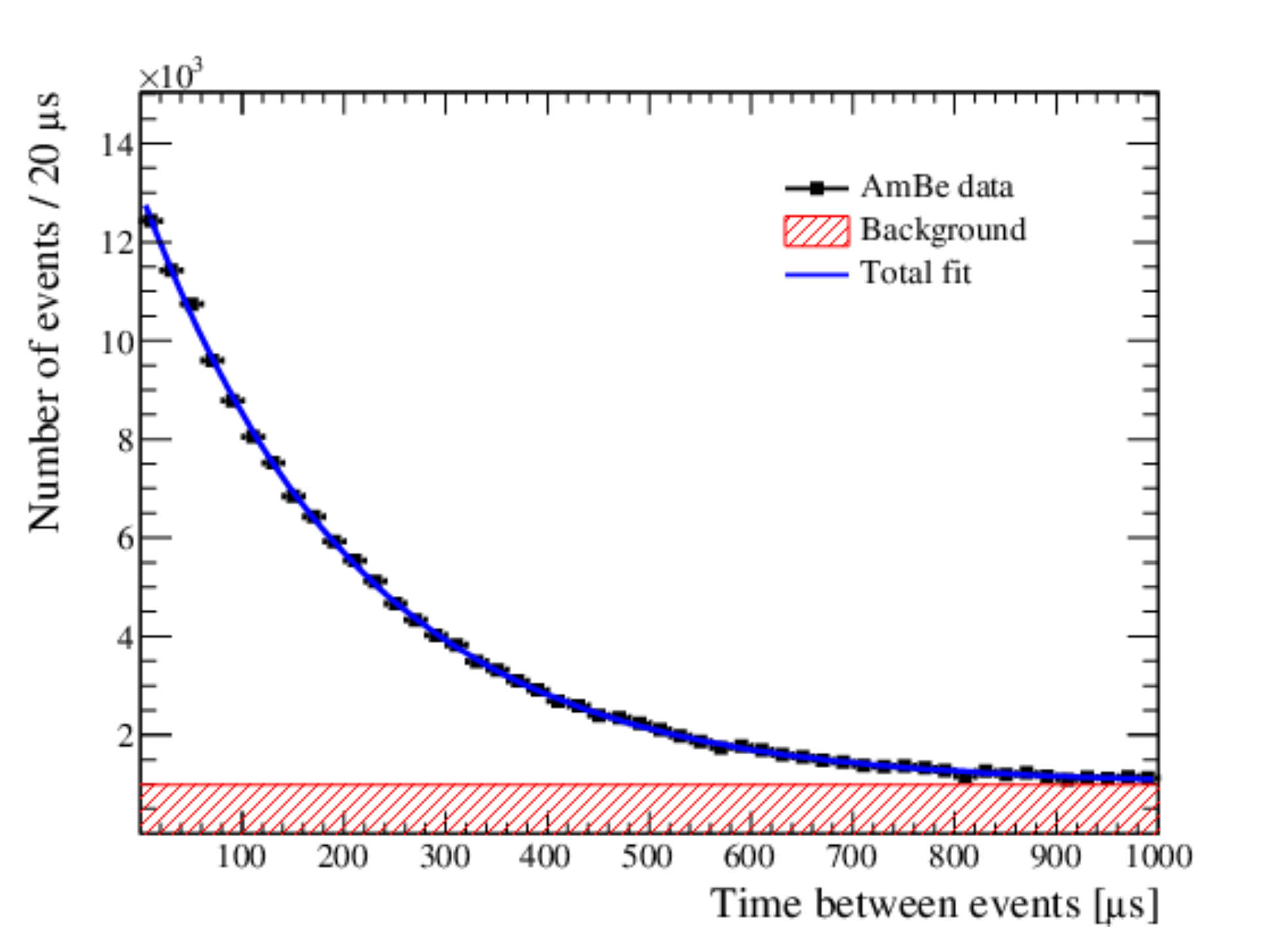}
 \caption{Time between events for an Am-Be source deployment at the detector center.  Coincidences are selected with the event proximity criterion (see text), fit with a single exponential plus a constant.}\label{fig:selectDT} 
\end{figure}

Figure~\ref{fig:selectDT} shows that the resulting time between events can be fitted with a single exponential plus a constant, due to the reduced background. Including statistical and systematic uncertainties, the fit yields a capture time constant of $209 \pm 3$~$\mu$s, consistent with the more precise one obtained in the primary analysis using Eq.~(\ref{eq:StatFit}), of $207.03 \pm 0.42$~$\mu$s.

The detection efficiency was calculated as the number of selected coincident neutrons determined from the fit divided by the total number of coincident neutrons available.  
The latter was estimated as the number of 4.4-MeV $\gamma$'s, which was determined by fitting a pure prompt spectrum and a background spectrum to the Am-Be data, and extracting the normalizations of the two spectra.  
The pure prompt spectrum was isolated in the Am-Be data using additional selection criteria: a reconstructed distance between events of $< $ 2.0~m, a maximum allowed time difference of 200~$\mu$s, and a more stringent requirement of at least 10 PMT hits occurring within 12~ns.  
The background spectrum was taken from data acquired immediately before and after the Am-Be calibration data.  

While suppressing the background by a factor of 3.5, this analysis achieved a detection efficiency just 2\% (absolute) lower than the efficiency obtained in the primary analysis, which required at least four PMT hits for a 2.2-MeV~$\gamma$.

\subsection{Systematic corrections and uncertainties}
\label{sec:syst}
Simulation showed that about 0.8\% of neutrons that coincide with a 4.4-MeV $\gamma$ will capture in the source encapsulation materials.  
Because the encapsulation is composed primarily of Delrin, which has a lower proton density than water, biases were induced in the measured efficiency and capture time constant.  
Corrections were derived by taking the difference between the fit results of simulations performed with and without the encapsulation.  
Uncertainties on these corrections were estimated by propagating the uncertainties in the measured density and dimensions of the Delrin.  
These corrections and uncertainties are tabulated for both the efficiency and capture time constant in Table~\ref{tab:syst}.  

The measured rate of events fluctuated due to transient variations in the trigger baseline, effectively changing the detector trigger threshold.  
The distribution of delayed-like event rate $R_d$ sampled in 1-second periods was found to be well described by a Maxwell-like distribution: $C(R_d-\mu)e^{-(R_d-\mu)^2/\sigma}$, where $C$, $\mu$, and $\sigma$ were fit for each run.  
Since $R_d$ is a constant in Eq.~(\ref{eq:StatFit}), a toy Monte Carlo model was used to evaluate the systematic effects of its fluctuation on both the capture time constant and the efficiency.  
Each event in a toy Monte Carlo dataset was assigned an ID (prompt $\gamma$, neutron, or background) and a trigger time.  
Events were generated using estimated values for the purity of 4.4-MeV $\gamma$'s, neutron detection efficiency, neutron capture time constant, and background rate, where the latter was sampled from the Maxwell-like distribution with $\mu$ and $\sigma$ set to their median values across all AV-internal runs.  
A correction to the fitted neutron capture time constant was calculated as the difference between the true Monte Carlo value and the fitted value.  This correction also addresses any bias in the fit for the time constant.  

To validate the correction, the capture time was evaluated as a function of $N_d$.  
As $N_d$ increases, the fluctuations in $R_d$ are suppressed, the Maxwell-like distribution becomes symmetric, and the correction vanishes.  Furthermore, the corrected capture time is consistent across all choices of $N_d$.  
An uncertainty for this systematic correction on the capture time was calculated by propagating the variation in $\mu$ and $\sigma$ across the AV-internal runs.
Similarly, the effect on the detection efficiency was obtained for the three central runs.
These corrections and uncertainties are tabulated in Table~\ref{tab:syst}.

\subsection{Results}
\label{sec:results}
Table~\ref{tab:syst} shows the fit results for Am-Be neutron detection efficiency and capture time constant in water, the systematic corrections from source encapsulation and background rate fluctuations and their uncertainties, and the corresponding final results.
The efficiency is the result obtained deploying the source at the center of the detector and the capture time constant is the average result from all the positions inside the AV.
The corrected capture time also includes two additional, minor uncertainties from temperature variation and neutron energy, which are described for the cross section calculation in Section~\ref{sec:CSsyst}.  

The efficiency to detect a neutron capture at the center of the SNO+ detector was $(49.08\pm0.39)\%$, using the normal trigger settings and data cleaning criteria of the SNO+ water phase.  
This efficiency for detecting 2.2-MeV $\gamma$'s is the highest among pure water Cherenkov detectors.  
The efficiency loss from the data cleaning criteria referred to in Section~\ref{sec:analysis} was evaluated as a function of the number of PMT hits, and then convolved with the constructed 2.2-MeV $\gamma$ hit distribution shown in Fig.~\ref{fig:NhitCenter}, yielding ($1.89\pm0.39$)\%.  

The neutron capture time constant was measured to be $\tau=202.35^{+0.87}_{-0.76}~\mu$s, similar to that of another large water Cherenkov detector (namely $202.6\pm3.7~\mu$s from Super-Kamiokande~\cite{Super-Kamiokande:2015xra}), but with significantly smaller uncertainties.

\begin{table}[tbp]\centering
  \caption{Summary of the measurements, systematic corrections, and uncertainties of the efficiency to detect a neutron capture at the center of the SNO+ detector, and of the capture time constant $\tau$.
Corrections are added to the fit results to obtain the final results.
See text for details.}\label{tab:syst}
\begin{tabular}{lcc}
\hline
\hline
& Efficiency~[\%] & $\tau$~[$\mu$s] \\
\hline
Fit result & $48.44\pm0.17$ & ~~$207.03\pm0.42$ \\
\hline
Source encapsulation & ~$0.43\pm0.20$ & $-2.86_{-0.54}^{+0.70}$ \\
Rate fluctuation & ~$0.21\pm0.29$ & $-1.78^{+0.23}_{-0.25}$ \\
\hline
Final result  & $49.08\pm0.39$ & $202.35^{+0.87}_{-0.76}$ \\
\hline
\hline
\end{tabular}
\end{table}

\section{Thermal neutron-proton capture cross section}
\label{sec:CS}

The capture time constant $\tau$ is converted to a thermal capture cross section $\sigma_{\mathrm{H},t}$ via 
\begin{equation}
\sigma_{\mathrm{H},t} = \frac{1}{\tau~v_{n,t}~n_\mathrm{H}},
\label{eq:crossSection}
\end{equation}
where $v_{n,t}$ is the thermal neutron velocity and $n_\mathrm{H}$ is the number density of hydrogen atoms.  

The typical value used for thermal neutron velocity $v_{n,t}$ is 2200~m/s, which corresponds to a kinetic energy of 0.02530~eV.  

The number density of hydrogen in the SNO+ detector was calculated as
\begin{equation}
 n_\mathrm{H} = \rho~w_\mathrm{H}~N_A~/~m_\mathrm{H},
\label{eq:AmBe:protonDensity}
\end{equation}
where $\rho$ is the density of water at the temperature and pressure at which the capture time was measured (0.9991$\times$10$^6$~g m$^{-3}$), $w_\mathrm{H}$ is the mass fraction of hydrogen in H$_2$O (11.19\%), $N_A$ is Avogadro's number (6.0221$\times$10$^{23}$~mol$^{-1}$), and $m_\mathrm{H}$ is the molar mass of hydrogen (1.0080~g~mol$^{-1}$).
These numbers give $n_\mathrm{H}$ = 0.6680$\times$10$^{29}$~m$^{-3}$.

\subsection{Systematic uncertainties}
\label{sec:CSsyst}
The temperature of the water affects both $v_n$ and $n_\mathrm{H}$.  
The number density $n_\mathrm{H}$ varies with temperature just as the density of water does, which is -0.015\%/$^\circ$C at 15$^\circ$C, the latter being the typical temperature during data acquisition.  
For hydrogen, the product of $v_n \sigma_\mathrm{H}(v_n)$ is extremely flat as a function of energy below $O$(10) keV, therefore little variation is expected.  This was quantified with Monte Carlo calculations using the energy-dependent cross section and a Maxwell$-$Boltzmann velocity distribution.  The product was found to change by -0.0022\%/$^\circ$C.  

A maximum difference of 2.4$^\circ$C was observed between the top and bottom of the volume of water beyond the PMTs.  Because there is no direct measurement of the water within the AV, a variation of 3$^\circ$ within the AV is assumed.  
Thus, the total systematic uncertainty from temperature is estimated to be 0.05\%, or 0.09~$\mu$s if applied to $\tau$.  

Since neutrons from an Am-Be source are emitted with MeV-scale energies, simulations were performed to evaluate any impact on the measurement of $\tau$ relative to using purely thermal neutrons.  
Simulations of thermal neutrons and of Am-Be neutrons at the center of the detector were analyzed following the event selections described in Section~\ref{sec:analysis}, and fitted with an exponential from 3~$\mu$s to 1000~$\mu$s, which resulted in indistinguishable time constants.  
A correction of ($-0.05\pm0.19$)~$\mu$s was added to $\sigma_{\mathrm{H},t}$, where the uncertainty reflects the precision of the simulations.

\subsection{Result}
\label{sec:CSresult}
The thermal capture time constant $\tau$ from Section~\ref{sec:results}, $v_{n,t}$, and $n_\mathrm{H}$ are combined via Eq.~(\ref{eq:crossSection}), yielding a thermal capture cross section of 
\begin{equation}
\sigma_{\mathrm{H},t} = 336.3^{+1.2}_{-1.5}~\mathrm{mb}, 
\label{eq:crossSectionResult}
\end{equation}
including the systematic uncertainties from temperature variation and the impact of the neutron energy spectrum.  

Dedicated experiments have measured the thermal neutron-proton capture cross section using strong-pulsed sources to create large numbers of neutrons in smaller water volumes.  The decay of these populations of neutrons was evaluated as a function of time.  
The most precise measurements are $334.2\pm0.5$~mb~\cite{Cox:1965} (1965) and $332.6\pm0.7$~mb~\cite{Cokinos:1977zz} (1977), followed by the result presented here. 
In contrast, the measurement presented here was made by analyzing the capture time distribution of individual neutrons, in a much larger, uniform, pure water Cherenkov detector.  The analysis considered the presence of a large random coincidence background, trigger threshold fluctuations, and the presence of the source container, as discussed in Sections~\ref{sec:minimalistic} and \ref{sec:syst}.  
In common with the other measurements, considerations were made for the energy spectra of source neutrons and variations in the temperature of the detector media, both of which were found to be small relative to the other uncertainties.

\section{Summary}

SNO+ collected data for nearly two years as a low-threshold water Cherenkov detector.  
The efficiency to detect 2.2-MeV $\gamma$'s was measured in a dedicated calibration campaign, and found to be 
centered around 50\% with a variation at the level of 1\% across the inner region of the detector.  
It was also found to be above 30\% outside the primary target volume, which if included in an analysis of uniformly-distributed signals such as reactor antineutrinos or supernova antineutrinos, would roughly double the fiducial volume.  
To our knowledge, these results establish SNO+ as the most efficient water Cherenkov detector for neutron captures on hydrogen.  
The neutron-hydrogen capture time constant was measured to be $202.35^{+0.87}_{-0.76}~\mu$s.  This was converted to a thermal neutron-proton capture cross section of $336.3^{+1.2}_{-1.5}$~mb.

\begin{acknowledgments}

Capital construction funds for the SNO+ 
experiment were provided by the Canada
Foundation for Innovation (CFI) and matching partners.  
This research was supported by:
{\bf Canada:}
Natural Sciences and Engineering Research Council,
the Canadian Institute for Advanced Research (CIFAR),
Queen's University at Kingston,
Ontario Ministry of Research, Innovation and Science,
Alberta Science and Research Investments Program,
National Research Council,
Federal Economic Development Initiative for Northern Ontario,
Northern Ontario Heritage Fund Corporation,
Ontario Early Researcher Awards,
the McDonald Institute;
{\bf US:}
Department of Energy Office of Nuclear Physics,
National Science Foundation,
the University of California, Berkeley,
Department of Energy National Nuclear Security Administration through the
Nuclear Science and Security Consortium;
{\bf UK:}
Science and Technology Facilities Council (STFC),
the European Union's Seventh Framework Programme under the European Research
Council (ERC) grant agreement,
the Marie Curie grant agreement;
{\bf Portugal:}
Funda\c{c}\~{a}o para a Ci\^{e}ncia e a Tecnologia (FCT-Portugal);
{\bf Germany:}
the Deutsche Forschungsgemeinschaft;
{\bf Mexico:}
DGAPA-UNAM and Consejo Nacional de Ciencia y Tecnolog\'{i}a.

We thank the SNO\raisebox{0.5ex}{\tiny\textbf{+}} technical staff for their strong contributions. We would
like to thank SNOLAB and its staff for support through underground space,
logistical and technical services. SNOLAB operations are supported by the
CFI and the Province of Ontario Ministry of
Research and Innovation, with underground access provided by Vale at the
Creighton mine site.

This research was enabled in part by support provided by WestGRID
(www.westgrid.ca) and Compute Canada (www.computecanada.ca) in particular
computer systems and support from the University of Alberta (www.ualberta.ca)
and from Simon Fraser University (www.sfu.ca) and by the GridPP Collaboration,
in particular computer systems and support from Rutherford Appleton Laboratory~\cite{Faulkner:2006px, Britton:2009ser}.  
Additional high-performance computing was provided
through the ``Illume'' cluster funded by the CFI
and Alberta Economic Development and Trade (EDT) and operated by
ComputeCanada and the Savio computational cluster resource provided by the
Berkeley Research Computing program at the University of California, Berkeley
(supported by the UC Berkeley Chancellor, Vice Chancellor for Research, and
Chief Information Officer). 
Additional long-term storage was provided by the
Fermilab Scientific Computing Division. Fermilab is managed by Fermi Research
Alliance, LLC (FRA) under Contract with the U.S. Department of Energy, Office
of Science, Office of High Energy Physics.
\end{acknowledgments}


\begin{thebibliography}{99}


\bibitem{Eguchi:2002dm}
  K.~Eguchi {\it et al.} [KamLAND Collaboration],
  Phys.\ Rev.\ Lett.\  {\bf 90}, 021802 (2003)
  doi:10.1103/PhysRevLett.90.021802
  [hep-ex/0212021].

\bibitem{Bellini:2010hy}
  G.~Bellini {\it et al.} [Borexino Collaboration],
  Phys.\ Lett.\ B {\bf 687}, 299 (2010)
  doi:10.1016/j.physletb.2010.03.051
  [arXiv:1003.0284 [hep-ex]].

\bibitem{Zhang:2013tua}
  H.~Zhang {\it et al.} [Super-Kamiokande Collaboration],
  Astropart.\ Phys.\  {\bf 60}, 41 (2015)
  doi:10.1016/j.astropartphys.2014.05.004
  [arXiv:1311.3738 [hep-ex]].


\bibitem{Boger:1999bb}
  J.~Boger {\it et al.} [SNO Collaboration],
  Nucl.\ Instrum.\ Meth.\ A {\bf 449}, 172 (2000)
  doi:10.1016/S0168-9002(99)01469-2
  [nucl-ex/9910016].

\bibitem{An:2015qga} 
  F.~P.~An {\it et al.} [Daya Bay Collaboration],
  Nucl.\ Instrum.\ Meth.\ A {\bf 811}, 133 (2016)
  doi:10.1016/j.nima.2015.11.144
  [arXiv:1508.03943 [physics.ins-det]].

\bibitem{Allemandou:2018vwb}
  N.~Allemandou {\it et al.} [STEREO Collaboration],
  JINST {\bf 13}, no. 07, P07009 (2018)
  doi:10.1088/1748-0221/13/07/P07009
  [arXiv:1804.09052 [physics.ins-det]].

\bibitem{Back:2019aqi} 
  A.~R.~Back {\it et al.} [ANNIE Collaboration],
  arXiv:1912.03186 [physics.ins-det].

\bibitem{Ashenfelter:2018zdm} 
  J.~Ashenfelter {\it et al.} [PROSPECT Collaboration],
  Nucl.\ Instrum.\ Meth.\ A {\bf 922}, 287 (2019)
  doi:10.1016/j.nima.2018.12.079
  [arXiv:1808.00097 [physics.ins-det]].

\bibitem{Agnes:2015qyz} 
  P.~Agnes {\it et al.} [DarkSide Collaboration],
  JINST {\bf 11}, no. 03, P03016 (2016)
  doi:10.1088/1748-0221/11/03/P03016
  [arXiv:1512.07896 [physics.ins-det]].


\bibitem{Andringa:2015tza}
  S.~Andringa {\it et al.} [SNO+ Collaboration],
  Adv.\ High Energy Phys.\  {\bf 2016}, 6194250 (2016)
  doi:10.1155/2016/6194250
  [arXiv:1508.05759 [physics.ins-det]].

\bibitem{Anderson:2018ukb}
  M.~Anderson {\it et al.} [SNO+ Collaboration],
  Phys.\ Rev.\ D {\bf 99}, no. 1, 012012 (2019)
  doi:10.1103/PhysRevD.99.012012
  [arXiv:1812.03355 [hep-ex]].

\bibitem{Anderson:2018byx}
  M.~Anderson {\it et al.} [SNO+ Collaboration],
  Phys.\ Rev.\ D {\bf 99}, no. 3, 032008 (2019)
  doi:10.1103/PhysRevD.99.032008
  [arXiv:1812.05552 [hep-ex]].


\bibitem{Loach:2008msa} 
  J.~C.~Loach, Ph.D. thesis, 2008, \url{https://www.sno.phy.queensu.ca/sno/papers/loach.pdf}


\bibitem{Delrin}
  Delrin is an acetal (polyoxymethylene) homopolymer from Dupont\texttrademark.



\bibitem{Super-Kamiokande:2015xra} 
  Y.~Zhang {\it et al.} [Super-Kamiokande Collaboration],
  Phys.\ Rev.\ D {\bf 93}, no. 1, 012004 (2016)
  doi:10.1103/PhysRevD.93.012004
  [arXiv:1509.08168 [hep-ex]].



\bibitem{Cox:1965}
  A.~E.~Cox, S.~A.~B.~Wynchank, and C.~H.~Collie,
  Nucl.\ Phys.\ {\bf 74}, 497 (1965).

\bibitem{Cokinos:1977zz} 
  D.~Cokinos and E.~Melkonian,
  Phys.\ Rev.\ C {\bf 15}, 1636 (1977).
  doi:10.1103/PhysRevC.15.1636



\bibitem{Faulkner:2006px} 
  P.~J.~W.~Faulkner {\it et al.} [GridPP Collaboration],
  J.\ Phys.\ G {\bf 32}, N1 (2006).
  doi:10.1088/0954-3899/32/1/N01

\bibitem{Britton:2009ser} 
  D.~Britton {\it et al.},
  Phil.\ Trans.\ Roy.\ Soc.\ Lond.\ A {\bf 367}, no. 1897, 2447 (2009).
  doi:10.1098/rsta.2009.0036

\end{thebibliography}
\end{document}